\documentclass[aip,jmp,graphicx,reprint]{revtex4-1}
\draft
\usepackage{dcolumn}
\usepackage{graphicx}
\usepackage{subfigure}
\begin{document}

\title{Enhancement of laser-driven electron acceleration in an ion channel}

\author{Alexey V. Arefiev}

\author{Vladimir N. Khudik}
\affiliation{Institute for Fusion Studies, The University of Texas, Austin, Texas 78712, USA}

\author{Marius Schollmeier}
\affiliation{Sandia National Laboratories, Albuquerque, New Mexico 87185, USA}

\date{\today}

\begin{abstract}

A long laser beam propagating through an underdense plasma produces a positively charged ion channel by expelling plasma electrons in the transverse direction. We consider the dynamics of a test electron in a resulting two-dimensional channel under the action of the laser field and the transverse electric field of the channel. A considerable enhancement of the axial momentum can be achieved in this case via amplification of betatron oscillations. It is shown that the oscillations can be parametrically amplified when the betatron frequency, which increases with the wave amplitude, becomes comparable to the frequency of its modulations. The modulations are caused by non-inertial (accelerated/decelerated) relativistic axial motion induced by the wave regardless of the angle between the laser electric field and the field of the channel. We have performed a parameter scan for a wide range of wave amplitudes and ion densities and we have found that, for a given density, there is a well pronounced wave amplitude threshold above which the maximum electron energy is considerably enhanced. We have also calculated a time-integrated electron spectrum produced by an ensemble of electrons with a spread in the initial transverse momentum. The numerical results show that the considerable energy enhancement is accompanied by spectrum broadening. The presented mechanism of energy enhancement is robust with respect to an axial increase of ion density, because it relies on a threshold phenomenon rather than on a narrow linear resonance.

\end{abstract}

\maketitle

\section{Introduction}

Novel applications that employ laser-target interactions continue to emerge with the ongoing improvement of high-power laser systems~\cite{HEDP2003}. It has been demonstrated that a laser-irradiated target in a properly chosen experimental setup can serve as a source of GeV electrons~\cite{Wang2013}, multi-MeV protons and ions~\cite{Fuchs2006,Flippo2010}, x-rays and gamma-rays~\cite{Kneip2008,Cipiccia2011}, and even copious positrons~\cite{Chen2013}. Despite the seeming diversity of the applications, there is a key unifying element: all of them require generation of energetic electrons by the laser. For that reason, the dynamics of electron heating and acceleration remains an active topic of research~\cite{Wilks1992,Pukhov1999,Kemp2012,Arefiev2012,Liu2013,Robinson2013}. 

The regime of laser interaction with the target electrons is greatly influenced by the density of the target. The interaction takes place only at the surface of the target if the density is over-critical, as in the case of solid density targets~\cite{Wilks1992,Kemp2012}. On the other hand, the laser beam is able to go through the target if the density is sub-critical, as in the case of gas jets
\cite{Kneip2009,Gahn2002,Mangles2005,Gahn1999,Walton2006}. Such low-density targets offer an extended interaction length with the laser, thus enabling a higher energy gain by the electrons. A sub-critical plasma layer can also naturally occur in experiments with solid-density targets where a significant pre-pulse is present. Our interest in electron heating is motivated by experiments with solid-density targets on fast proton generation using pulses with duration of 1 to 5 ps~\cite{Flippo2010}. A prepulse in these experiments can create a transparent preplasma, extending many wavelengths from the target surface along the beam path. The main pulse then interacts with a low-density plasma before reaching the target. It is important to understand if such interactions can generate hot electrons in addition to the ones produced at the critical surface~\cite{Wilks1992,Kemp2012}. 

In order to address this issue, one has to consider laser interaction with a low-density sub-critical plasma. It is well known that the electron acceleration mechanism in this regime depends significantly on laser pulse duration. A laser-produced wakefield plays the dominant role when the beam is relatively short, with the duration shorter or comparable to the plasma wave period~\cite{Esarey2009}. However, the wakefield mechanism becomes less efficient for longer laser pulses~\cite{Malka2001,Mangles2005}. This would be the case for typical pulse durations used in experiments on fast proton generation. In the case of a long laser beam, its ponderomotive pressure tends to expel plasma electrons from the beam in the transverse direction~\cite{Willingale2013,Iwata2013}. The plasma mitigates this expulsion by generating a counteracting electric field via charge separation. As a result, the laser can create a positively charged channel with a quasi-static transverse electric field evolving on an ion time scale. Such channels and the corresponding transverse electric field are routinely observed in simulations of laser interactions with underdense plasmas~\cite{Sentoku2006,Li2008,Sarri2010,Friou2012}.

An electron accelerated by the laser can become confined in the channel, oscillating across the channel while moving along with the beam. Considerable acceleration of electrons has been observed experimentally in such channels~\cite{Mangles2005, Kneip2009,Gahn2002,Gahn1999, Kitagawa2004, Kneip2008, Walton2006}. Although the plasma in these experiments was significantly underdense, the measured electron energies were much greater than the energy expected in a vacuum for a laser beam of the same intensity. The electric field of the channel clearly plays a critical role in enhancing the electron energy gain during its interaction with the laser pulse. 

Several mechanisms that involve the transverse~\cite{Pukhov1999,Arefiev2012,Arefiev2012b} or axial~\cite{Robinson2013} field of the channel have been proposed as possible explanations for the enhanced electron energy gain. The region with a considerable axial field is typically located near the channel opening. Its axial extend is much shorter than the length of the channel. The effect of the axial field can therefore be accounted for through initial conditions when analyzing electron dynamics over the length of the channel~\cite{Robinson2013}. The focus of this work is on the role played by the transverse electrostatic field of the channel in enhancing the electron energy gain. 

In what follows, we consider the dynamics of an electron irradiated by a linearly polarized electromagnetic wave in a  two-dimensional steady-state ion channel. The main advantage of the two-dimensional setup is that for a given laser amplitude the amplitude of the laser electric field that directly drives the betatron oscillations can be changed by changing the laser polarization angle (the angle between the laser electric field and the field of the channel). This allows us to examine the interplay between the electrostatic field of the channel and the oscillating electric field of the laser. It is shown that a considerable enhancement of the electron energy, which is predominantly associated with the longitudinal motion, requires a significant amplification of the betatron oscillations. 

Our qualitative analysis indicates that there are two factors contributing to the amplification of the betatron oscillations: (1) the betatron oscillations can become parametrically unstable~\cite{Arefiev2012}, which would cause their amplitude to grow, and (2) the betatron oscillations can be directly driven by the laser field. The role of the two factors can be distinguished because the parametric instability is caused by modulations of the betatron frequency and it can develop in the absence of a driving electric field. The origin of these modulations is the non-inertial (accelerated/decelerated) relativistic axial motion induced by the linearly polarized wave. We find that the amplification of the transverse oscillations and the subsequent enhancement of the axial momentum occur regardless of the laser polarization. 
This result points to the key role played by the parametric amplification. We also find that the enhancement factor  increases when the laser electric field and the field of the channel are collinear. It is plausible that this increase is 
linked to a nonlinear betatron resonance~\cite{Pukhov1999}, because the betatron frequency becomes comparable to the frequency of the driving field for the same range of parameters where the condition for the parametric resonance is satisfied.

The rest of the paper is organized as follows. In Sec.~\ref{Sec_1}, we formulate a single electron model for an electron irradiated by an incoming electromagnetic wave in a steady-state two-dimensional ion channel. In Sec.~\ref{Sec_vac}, we summarize the key features of electron acceleration in vacuum to establish the context for the qualitative analysis of electron motion in the channel of Secs.~\ref{Sec_2}. Section \ref{Sec_2_amp} provides a more detailed analysis of the onset of the parametric instability. The qualitative analysis of Secs.~\ref{Sec_2} is illustrated by a number of numerical examples in Sec.~\ref{Sec_5}. In Sec.~\ref{Sec_6}, we calculate the threshold for the enhancement of the electron energy and the enhancement factor. Sec.~\ref{Sec_7} presents electron spectra for an ensemble of electrons with a spread in the initial transverse momentum in the regimes with and without the enhancement of the electron energy. Section~\ref{sum} summarizes our findings.


\section{Single electron model} \label{Sec_1}

\begin{figure}
  \includegraphics[width=0.9\columnwidth]{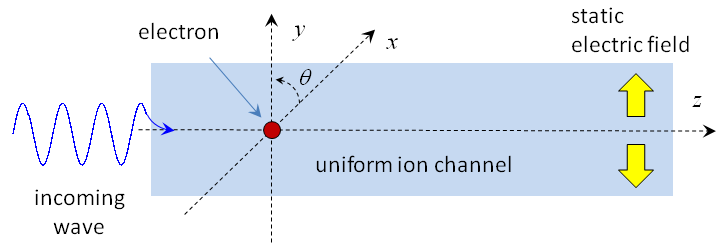}
  \caption{Schematic setup of the single-electron model.} \label{Fig3}
\end{figure}

We consider the dynamics of a single electron placed in an ion channel and irradiated by a plane electromagnetic wave. The main simplification in considering a single particle is that all the fields are given. We use the setup shown in Fig.~\ref{Fig3}, with a Cartesian system of coordinates $(x,y,z)$. The channel is effectively a slab of immobile ions with density $n_0$. The $(x,z)$-plane at $y=0$ is the midplane of the ion slab, with the $y$-axis directed across the slab. The ions create a static electric field directed only along the $y$-axis that vanishes at $y=0$. The plane electromagnetic wave propagates in the positive direction along the $z$ axis. The direction of the laser electric field is specified by the polarization angle $\theta$. In this case, the wave field is described by a normalized vector potential
\begin{equation}
{\bf{a}} (z,t) = a(\xi) \left[ {\bf{e}}_x \cos \theta   + {\bf{e}}_y \sin \theta \right]
\end{equation}
that is only a function of a dimensionless variable
\begin{equation} \label{Eq_xi}
	\xi \equiv \omega (t - z / c) ,
\end{equation}
where $\omega$ is the wave frequency, $c$ is the speed of light, $t$ is the time in the channel frame of reference, and ${\bf{e}}_x$ and ${\bf{e}}_y$ are unit vectors. In the analysis that follows this section, we consider pulses with
\begin{eqnarray} \label{pulse_shape}
&& a(\xi) = a_*(\xi) \sin(\xi), \\
&& 0 \leq a_* \leq a_0,
\end{eqnarray}
where $a_*(\xi)$ is a given slowly varying envelope. 

The equations that govern electron motion in the channel are derived in Appendix \ref{Ap-1} using a Hamiltonian for an electron irradiated by an electromagnetic wave in a given electrostatic field. The equations for the motion in the $(y,z)$-plane are 
\begin{eqnarray}
&& \frac{d }{d \tau} \left( \frac{p_y}{m_e c} - a \sin \theta \right) = - \gamma \frac{\omega_p^2}{\omega^2}  \frac{\omega}{c} y, \label{Eq1} \\
&& \frac{d}{d \tau} \left( \frac{p_z}{m_e c} \right)  = \left( \frac{p_y}{m_e c} - a \sin \theta \right) \sin \theta \frac{d a}{d \xi} + \frac{d}{d \xi} \left( \frac{a^2}{2} \right) , \label{Eq2}\\
&& \frac{d}{d \tau} \left( \frac{\omega}{c} y \right) =  \frac{p_y}{m_e c} , \label{Eq3}\\
&& \frac{d \xi}{d \tau} =  \gamma - \frac{p_z}{m_e c}, \label{Eq4}
\end{eqnarray}
where $p_y$ and $p_z$ are components of the electron momentum,
\begin{equation}
\gamma = \sqrt{1 + a^2 \cos^2 \theta + \left(p_y/m_e c\right)^2 + \left(p_z / m_e c \right)^2} \label{Eq5}
\end{equation}
is the relativistic factor, and $\tau$ is a dimensionless proper time defined by the relation 
\begin{equation} \label{Eq6}
	d \tau /d t = \omega / \gamma.
\end{equation}
Here $\omega_p \equiv \sqrt{4 \pi n_0 e^2/m_e}$ is the plasma frequency, where $e$ and $m_e$ are the electron charge and mass. Equations~(\ref{Eq1}) and (\ref{Eq2}) are transverse and parallel momentum balance equations, whereas Eqs.~(\ref{Eq3}) and (\ref{Eq4}) relate the time evolution of the corresponding transverse and axial coordinates. These equations explicitly take into account that $p_x - m_e c a \cos \theta = 0$ in the considered set-up.

If the static electric field created by the ions is independent of the longitudinal coordinate $z$, as in the case of the channel shown in Fig.~\ref{Fig3}, then Eqs.~(\ref{Eq1})~- (\ref{Eq4}) have the following integral of motion:
\begin{equation} \label{R_main_0}
I \equiv \gamma - \frac{p_z}{m_e c} + \frac{\omega_p^2}{c^2} \frac{y^2}{2}.
\end{equation}
For an electron with initial axial momentum $p_0$ and initial transverse displacement $y_0$, we have
\begin{equation} \label{R_main}
I = \sqrt{1 + \frac{p_0^2}{m_e^2 c^2}} - \frac{p_0}{m_e c} + \frac{\omega_p^2}{c^2} \frac{y_0^2}{2}.
\end{equation}
It should be noted that $I$ is always positive. Equation~(\ref{Eq2}) can be replaced with Eq.~(\ref{R_main_0}). Using this equation, we can eliminate $p_z$ in Eq.~(\ref{Eq4}) and in the expression for $\gamma$ given by Eq.~(\ref{Eq5}). We next combine Eqs.~(\ref{Eq1}) and (\ref{Eq3}) to obtain two coupled equations for $y$ and $\xi$:
\begin{eqnarray}
&& \frac{d^2 y}{d \tau^2} + \gamma \frac{\omega_p^2}{\omega^2}  y =  \frac{c}{\omega} \frac{d \xi}{d \tau} \frac{d a}{d \xi} \sin \theta, \label{main1} \\
&& \frac{d \xi}{d \tau}  =  I - \frac{1}{2} \frac{\omega_p^2}{c^2}  y^2, \label{main2}
\end{eqnarray}
where the relativistic $\gamma$-factor is now given by
\begin{equation}
\gamma = \frac{1}{2} \left. \left[ 1 + a^2 \cos^2 \theta + \frac{\omega^2}{c^2} \left( \frac{dy}{d \tau} \right)^2 +  \left( \frac{d \xi}{d \tau} \right)^2 \right] \right/ \frac{d \xi}{d \tau} . \label{main3}
\end{equation}
Equations (\ref{main1}) - (\ref{main3}) are equivalent to Eqs.~(\ref{Eq1})~- (\ref{Eq5}) and we will use them instead when analyzing electron oscillations across the channel. Equation (\ref{main2}) can be interpreted as the axial dephasing, as it gives the rate at which the phase of the wave field sampled by the electron changes in an instantaneous inertial frame where the electron is at rest. Once the electron trajectory is found by solving Eqs.~(\ref{main1}) and (\ref{main2}), the components of the electron momentum can be calculated using the expressions
\begin{eqnarray}
&& p_x =  m_e c a \cos \theta, \label{main4}\\
&& p_y = m_e \omega dy/d\tau, \label{main5}\\
&& p_z = \left( \gamma + \frac{\omega_p^2}{c^2} \frac{y^2}{2} - I \right) m_e c \label{main6}
\end{eqnarray}
and time $t$ can be calculated by integrating Eq.~(\ref{Eq6}) with $\gamma$ given by Eq.~(\ref{main3}).


\section{Electron dynamics in vacuum} \label{Sec_vac}

In this section, we review the key features of electron acceleration in vacuum to establish the context for the 
later analysis of electron motion in an ion channel. The results for the vacuum regime follow from the already derived equations by setting $\omega_p = 0$. Specifically, it follows from Eq.~(\ref{main2}) that $d \xi / d \tau$ is a constant, $d \xi / d \tau = I$. Since $I$ is always positive, $\xi$ will monotonically increase with time. We assume that the electron interacts with the laser at $t>0$ and that it is initially ($t=0$) located at $y=0$ and $z=0$, such that initially $\xi =0$.

In order to find $dy/d\tau$ from Eq.~(\ref{main1}), we rewrite the first term as 
$d^2y/d\tau^2 = (d \xi/ d \tau) d(d y/d\tau)/d \xi$ and set $\omega_p = 0$, effectively removing the second term on the left-hand side. The resulting equation is easily integrated over $\xi$, which yields $dy/d\tau = c a \sin \theta /\omega$ for the case of $dy/d\tau = 0$ and $a = 0$ at $\xi = 0$. Taking into account that $d \xi / d \tau = I$ and, therefore, $dy/d\tau = I d y/d \xi$, we can integrate the equation over $\xi$ one more time to find that
\begin{equation} \label{sol-1_0}
y(\xi) = \frac{c \sin \theta}{\omega I}  \int_0^{\xi}   a(\xi') d \xi'.
\end{equation}
For a pulse with $a = a_*(\xi) \sin(\xi)$ [see Eq.~(\ref{pulse_shape})], we have 
\begin{equation} \label{sol-1_0_2}
y(\xi) = - \frac{c \sin \theta}{\omega I}  a_*(\xi) \cos(\xi) .
\end{equation}

We next use the expressions for $d \xi / d \tau$ and $dy/d\tau$ to find from Eqs.~(\ref{main3})~-~(\ref{main6}) the $\gamma$-factor and the electron momentum:
\begin{eqnarray} 
&& \gamma = \left. \left( 1 + a^2 + I^2 \right) \right/ 2 I, \label{main3_vac} \\
&& p_x =  m_e c a \cos \theta, \label{main4_vac}\\
&& p_y = m_e c a \sin \theta, \label{main5_vac}\\
&& p_z = m_e c \left. \left( 1 + a^2 - I^2 \right) \right/ 2 I, \label{pzA2}
\end{eqnarray}
where Eq.~(\ref{pzA2}) is derived by using expression~(\ref{main3_vac}) for $\gamma$ and setting  $\omega_p = 0$. According to Eq.~(\ref{main3_vac}), the maximum $\gamma$-factor that the electron can achieve accelerating in a vacuum is
\begin{equation} \label{gamma_vac}
\gamma_{vac} \equiv \left. \left( 1 + a_0^2 + I^2 \right) \right/ 2 I,
\end{equation}
where $a_0 \equiv \max a_*$ is the maximum wave amplitude. In the case of an electron that is initially at rest $(p_0 = 0)$, we have $I=1$ and $\gamma_{vac} = 1 + \left. a_0^2 \right/ 2$. The axial component of the momentum becomes dominant at ultrarelativistic wave amplitudes ($a \gg 1$), since $p_z \left/ \sqrt{p_x^2 + p_y^2} \right. = |a| /2$. If the electron is moving forward with relativistic axial momentum prior to the interaction with the laser ($p_0 / m_e c \gg 1$), then $I \approx m_e c / 2 p_0 \ll 1$ and, as a result, $\gamma \approx a_0^2 (p_0 / m_e c)$. Therefore, the electron $\gamma$-factor for a given wave amplitude increases proportionally to the initial $\gamma$-factor of an initially relativistic electron [$\gamma(t=0) \approx p_0 / m_e c \approx 1/2I$].


\section{Qualitative analysis of\\electron motion in the ion channel} \label{Sec_2}

We are considering a channel whose density is significantly subcritical, such that $\omega_p / \omega \ll 1$. In the limit of very low density, $\omega_p / \omega \rightarrow 0$, the electron motion is described by the vacuum solution derived in Sec.~\ref{Sec_vac}. We are looking for the regimes where ultrarelativistic electron motion in a wave with $a \gg 1$ can lead to a considerable enhancement of the electron axial momentum compared to the vacuum case. Since the static electric field has no axial component, an enhancement of the axial acceleration can only result from changes in electron oscillations across the channel (along the $y$-axis). 

The integral of motion $I$ given by Eq.~(\ref{R_main_0}) suggests one method for achieving an enhancement of the axial momentum. If the amplitude of the transverse oscillations is amplified and approaches 
\begin{eqnarray} \label{y_crit}
&& y_* \equiv \left. \sqrt{2I} c \right/ \omega_p,
\end{eqnarray}
then $\gamma - p_z/m_e c$ becomes vanishingly small according to Eq.~(\ref{R_main_0}). The combination $\gamma - p_z/m_e c$ is incidentally the dephasing rate and it can only vanish if $p_z \rightarrow +\infty$. We therefore conclude that one way to enhance $p_z$ is to amplify the transverse oscillations. 

In order to determine the conditions required for the amplification, we consider an electron with a small initial displacement across the channel, $|y_0| \ll y_*$, irradiated by a pulse with $a = a_*(\xi) \sin(\xi)$, where $a_*$ is a slowly changing envelope that increases from 0 to $a_0 \gg 1$. At the leading edge of the pulse, the amplitude of the transverse oscillations remains small, $|y| \ll y_*$. The axial dephasing rate and, as a result, the $\gamma$-factor are then the same as in the vacuum regime, with $d\xi/d \tau \approx I$ and $\gamma$ given by Eq.~(\ref{main3_vac}). Using these expressions in Eq.~(\ref{main1}) for the transverse oscillations, we find that  
\begin{eqnarray} 
 \frac{d^2 y}{d \xi^2} &&+ \left[ \frac{1 + I^2 + a_*^2(\xi)/2}{2I^3} - \frac{a_*^2(\xi)}{4I^3} \cos(2\xi) \right] \frac{\omega_p^2}{\omega^2}  y \nonumber \\
&& =  \frac{c}{\omega I} \sin \theta a_*(\xi) \cos(\xi), \label{main1_2}
\end{eqnarray}
where we took into account that  $d^2y/d\tau^2 = I^2 d^2y/d\xi^2$ and $d a/d \xi \approx a_* \cos(\xi)$. 

Equation~(\ref{main1_2}) is similar to that of an oscillator with a modulated natural frequency driven by an external force. The amplitude of the natural frequency is
\begin{equation}
\Omega \equiv \sqrt{\frac{1 + I^2 + a_*^2(\xi)}{2I^3}} \frac{\omega_p}{\omega},
\end{equation}
whereas the frequency of the external force and the frequency of the modulations are 1 and 2, respectively. We assume that the natural frequency is small at non-relativistic wave intensities due to the low ion density, i.e., 
$\Omega \ll 1$ for $a_* \ll 1$. This condition limits the initial axial momentum:
\begin{equation}
\frac{p_0}{m_e c} \ll \left( \frac{\omega}{2 \omega_p} \right)^{2/3},
\end{equation}
where we took into account that $I \approx m_e c/2 p_0$ for $p_0 \gg m_e c$.

Under our assumptions, the natural frequency is small at the leading edge of the laser pulse ($a_* \leq 1$). This frequency mismatch $(\Omega \ll 1)$ means that a resonant interaction that can lead to an amplification of the transverse oscillations is not possible. As the amplitude of the field sampled by the electron increases and becomes ultrarelativistic ($a_* \gg 1$), $\Omega$ starts to grow proportionally to $a_*$. The natural frequency $\Omega$ becomes comparable to the frequency of its own modulations and to the frequency of the external force in Eq.~(\ref{main1_2}) at $a_* \approx a_{cr}$, where
\begin{equation} \label{crit_amplitude}
a_{cr} = \sqrt{2} I^{3/2} \omega/\omega_p
\end{equation}
is defined by the condition $\Omega = 1$. In what follows, we consider how this frequency match affects the transverse oscillations and axial acceleration in two limiting cases: a) the wave electric field is directed along the $x$-axis ($\theta = 0$), so that the external force in Eq.~(\ref{main1_2}) vanishes; b) the wave electric field is directed along the $y$-axis ($\theta = \pi/2$), so that the external force directly drives the oscillations.

At $\theta=0$, Eq.~(\ref{main1_2}) describes a parametric oscillator without an external driving force. Natural oscillations in this system are stable if their frequency is considerably less than the frequency of the modulations $(\Omega \ll 1)$. There is a frequency threshold that is roughly comparable to the frequency of the modulations ($\Omega \approx 1$) above which the oscillations become parametrically unstable and their amplitude grows exponentially. However, the amplification of the transverse oscillations causes an enhancement of the axial acceleration only after $|y|$ becomes comparable to $y_*$. Therefore, the electron dynamics in a pulse with a gradually increasing envelope has three subsequent stages: 1) stable oscillations at $a_* \ll a_{cr}$ ($\Omega \ll 1$) with $|y| \ll y_*$; 2) exponential growth of the oscillations at $a_* \approx a_{cr}$ ($\Omega \ll 1$) while $|y| \ll y_*$; 3) enhancement of the axial acceleration that occurs once $|y|$ becomes comparable to $y_*$ as a result of the exponential growth. The linear equation for the transverse oscillations~(\ref{main1_2}) remains valid during the parametric amplification while $|y| \ll y_*$ and it can therefore be used to quantitatively determine the threshold and the exponential growth rate for the parametric amplification, as done in Sec.~\ref{Sec_2_amp}.

In order to estimate when the enhancement of the axial acceleration occurs after the onset of the parametric amplification, we assume that the exponential growth rate $\kappa$ is given [$y \propto \exp(\kappa \xi)$]. The change of phase needed for the amplitude to increase from $y_0$ to $y_*$ is $\Delta \xi \approx \kappa^{-1} \ln(y_*/y_0)$. This corresponds to $\Delta \tau \approx I^{-1} \Delta \xi$, since $d \xi / d \tau \approx I$. The time that elapses during the amplification is $\Delta t \approx \gamma I^{-1} \Delta \xi / \omega$ according to Eq.~(\ref{Eq6}). The electron is moving axially with the velocity close to the speed of light and, therefore, it travels a distance $\Delta z \approx c \Delta t$ before the amplitude of the transverse oscillations reaches $y_*$. The axial motion is essentially the same as in the vacuum case while $|y| \ll y_*$, so that $\gamma \approx a_0^2 / 2 I$ [see Eq.~(\ref{main3_vac})]. We combine these estimates to obtain
\begin{equation}
\Delta z \approx \frac{c}{\omega} \frac{a_0^2}{2 \kappa I^2} \ln(y_*/y_0).
\end{equation}
The length of the plasma channel has to exceed $\Delta z$ in order for the parametric instability to cause an enhancement of the axial momentum that occurs only after the amplitude of the transverse oscillations becomes comparable to $y_*$. A detailed analysis of the growth rate $\kappa$ is given in Sec.~\ref{Sec_2_amp}.

At $\theta=\pi/2$, the laser electric field directly drives electron oscillations across the channel. The corresponding driving force in Eq.~(\ref{main1_2}) causes the amplitude of the transverse oscillations to grow with the increase of the wave envelope $a_*$. If $\Omega \ll 1$ ($a_* \ll a_{cr}$), then the second term in Eq.~(\ref{main1_2}) can be neglected compared to the first term, because the frequency of the driven oscillations is equal to unity, which is the frequency of the driving force. The resulting equation is the same as in the vacuum case and hence $y(\xi)$ is given by Eq.~(\ref{sol-1_0_2}). According to this equation, the amplitude of the oscillations becomes comparable to $y_*$ at $a_* \approx a_{cr}$, which corresponds to $\Omega \approx 1$. We thus conclude that the transverse and longitudinal electron motion become influenced by the static field of the channel simultaneously as $a_*$ approaches $a_{cr}$, whereas the electron essentially moves as a free electron while $a_* \ll a_{cr}$. In contrast to the case of $\theta=0$, there is no intermediate regime at $a_* \approx a_{cr}$ where the axial motion can still be treated as the vacuum motion.  

The linear equation~(\ref{main1_2}) loses its applicability as $a_*$ approaches $a_{cr}$ at $\theta=\pi/2$ because $|y|$ can no longer be treated as small. However, the fact that 
the natural frequency approaches both the frequency of the modulations and the frequency of the driving force at the limit of the applicability suggests that the oscillations might become unstable in the nonlinear regime. Numerical solutions presented in Sec.~\ref{Sec_5} confirm that this is indeed the case. 
The electron dynamics for an intermediate range of polarization angles that are not very small ($\theta \sim 1$) is similar to the case of $\theta = \pi/2$. We show in Secs.~\ref{Sec_5} and \ref{Sec_6} that there is a well pronounced wave amplitude threshold close to $a_{cr}$ in the nonlinear regime. The transverse oscillations become unstable when $a_*$ exceeds the threshold and their amplitude $|y|$ approaches $y_*$, which causes a considerable enhancement of the axial acceleration.


\begin{figure}[tb]
  \centering
  \subfigure{\includegraphics[scale=0.5]{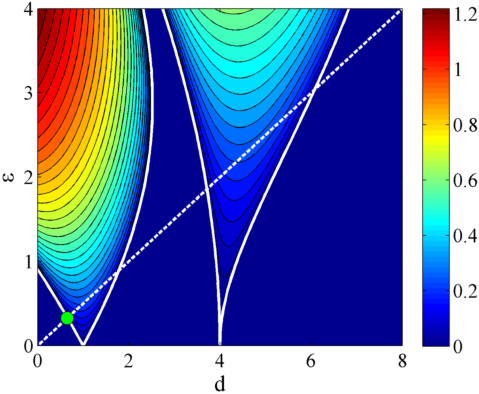} } \\
  \subfigure{\includegraphics[scale=0.45]{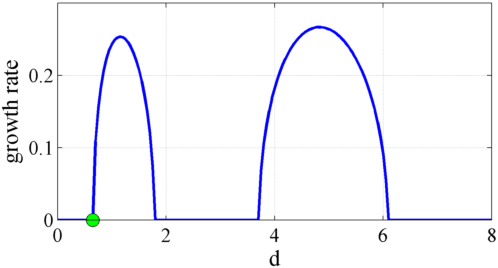}} 
\caption{The growth-rate $\nu$ for unstable solutions of Eq.~(\ref{Eq_M}). The solid curves in the upper plot are the boundaries of stable regions where $\nu = 0$. The lower plot shows the growth rate along the dashed line in the upper plot ($\epsilon = d/2$).}
\label{fig:stability}
\end{figure}

\section{Parametric amplification of betatron oscillations} \label{Sec_2_amp}

In this section we consider in more detail the onset of the parametric instability in the case of a pulse that has no electric field component across the channel ($\theta = 0$).  

As shown in Sec.~\ref{Sec_2}, transverse electron oscillations across the channel evolve according to  Eq.~(\ref{main1_2}) before their amplitude $y$ becomes comparable to $y_*$. In order to calculate the threshold for the onset of the parametric instability, we assume that the pulse envelope remains constant, $a_*(\xi) = a_0$, following an initial stage where it increases from $0$ to $a_0$. We therefore set $a_*(\xi) = a_0$ and $\theta = 0$ in Eq.~(\ref{main1_2}), which reduces it to the Mathieu equation~\cite{Ince,McLachlan} with two free parameters:
\begin{eqnarray} \label{Eq_M}
&& \frac{d^2 y}{d \xi^2} + \left[ d - 2 \epsilon \cos (2\xi) \right]  y =  0, \\
&& d \equiv 2 \epsilon \left[ \frac{2 (1 + I^2)}{a_0^2} + 1 \right], \label{b}\\
&& \epsilon \equiv \frac{a_0^2}{8 I^3} \frac{\omega_p^2}{\omega^2}.
\end{eqnarray}
Equation~(\ref{Eq_M}) can have an exponentially growing solution~\cite{McLachlan} $y(\xi) \propto \exp(\nu \xi)$, where the growth rate $\nu$ depends on $d$ and $\epsilon$. The upper plot of Fig.~\ref{fig:stability} shows $\nu$ 
for a wide range of parameters. The solid curves are the boundaries of stable regions where $\nu = 0$. 

In the case of a wave with an ultrarelativistic amplitude ($a_0 \gg 1$), the stability of the solutions is effectively determined by a single parameter, because $d \approx 2 \epsilon$. The dashed line in the upper plot of Fig.~\ref{fig:stability} corresponds to $d = 2 \epsilon$ and the curve in the lower plot of Fig.~\ref{fig:stability} gives the growth rate along this line. The parameter $\epsilon$ is related to the amplitude of the natural frequency $\Omega$ introduced in Sec.~\ref{Sec_2}, with $\Omega = 2 \sqrt{\epsilon}$ for $a_* = a_0 \gg 1$. The regime with $\Omega \ll 1$ thus corresponds to the stable region near the origin of the stability diagram ($\epsilon \ll 1$ and $d \ll 1$). The instability threshold is located at $d \approx 0.66$ for the case of $d = 2 \epsilon$, which corresponds to $\Omega \approx 1.15$. It is shown with a circle in both plots of Fig.~\ref{fig:stability}. Therefore, we conclude that the combination $a_0 \omega_p / \omega$ has to exceed 
\begin{equation} \label{thr2}
a_0 \omega_p / \omega \approx 1.62 I^{3/2}
\end{equation}
in order for the transverse oscillations to be parametrically amplified. It follows from Eq.~(\ref{thr2}) that the parametric amplification is feasible even in a significantly underdense channel ($\omega \gg \omega_p$) if the wave amplitude is sufficiently high. 

It is important to point out that the betatron oscillations will be amplified as long as $d$ exceeds the threshold value ($d \approx 0.66$) even if the corresponding growth rate according to Fig.~\ref{fig:stability} vanishes. For example, if $d = 3$, then the growth rate is zero. However, $d$ increases slowly from 0 to 3 because of the slowly increasing laser pulse envelope. Once $d$ exceeds the threshold value, the transverse oscillations begin to grow exponentially. They will quickly become nonlinear before the growth rate can vanish due to the continuous increase of $d$.


In order to provide a qualitative explanation for the dependence of the parametric amplification threshold on wave amplitude $a_0$, let us consider how the electron motion across the channel changes at ultrarelativistic wave intensities. For simplicity, we limit our consideration to the case of an electron that was initially at rest $(I=1)$ and that is irradiated by a linearly polarized wave that has no electric field component across the channel. At $a_0 \gg 1$, the electron is pushed primarily forward by the wave and it is moving in the axial direction with $\gamma \approx a_0^2/2$. In the reference frame of the channel, the ion density is significantly underdense, with $\omega_p \ll \omega$. However, the ion density perceived by the electron is enhanced by a factor of $\gamma$. This enhancement is a result of the length contraction in an instantaneous inertial reference frame co-moving axially with the electron. In the case of a linearly polarized wave, the electron experiences alternating axial acceleration and deceleration caused by axial gradients of the wave pressure. Therefore, the axial motion induced by the laser field not only significantly enhances, but also modulates the perceived ion density and, consequently, the restoring force from the ions. Another way of looking at this is by noting that a static transverse electric field is enhanced by a factor of $\gamma$ in the instantaneous reference frame co-moving with the electron along the axis of the channel. This fact is automatically accounted for in Eq.~(\ref{main1}) for the transverse electron motion, where the second term representing the restoring force is proportional to the $\gamma$-factor.

\begin{figure}[tb]
  \centering
  \subfigure{\includegraphics[scale=0.55]{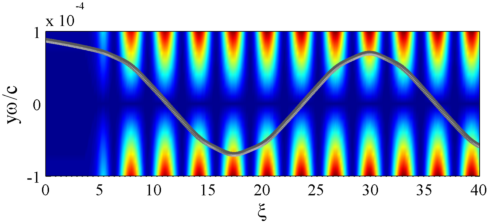} } \\
  \subfigure{\includegraphics[scale=0.55]{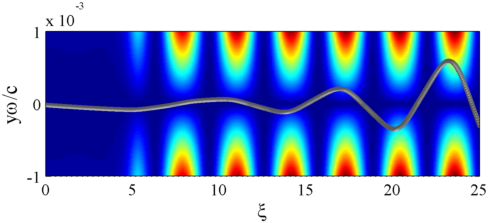}} 
\caption{Electron trajectory and color-coded amplitude of the restoring force in stable (upper panel) and unstable (lower panel) regimes. The peak field amplitude is $a_0 = 10$ in both cases, whereas $\omega_p/\omega = 0.05$ in the upper panel and $\omega_p/\omega = 0.175$ in the lower panel.}
\label{fig:traj}
\end{figure}

If $\gamma \omega_p \ll \omega$, then the restoring force rapidly oscillates as the electron slowly traverses the channel. An example of electron motion in this regime is shown in Fig.~\ref{fig:traj} (upper panel), where the trajectory is a solution of Eq.~(\ref{main1_2}) without the driving force on the right-hand side. The color-coding represents $| \gamma (\omega_p^2 / \omega^2) y |$ as a function of $\xi$ and $y$. At the electron location, this quantity is proportional to the restoring force acting on the electron. The electron trajectory remains stable, because the modulations are too rapid compared to the transverse oscillations and the jitter of the restoring force averages out during the slow transverse motion. The restoring force increases with the increase of the laser intensity at a given ion density, which leads to more rapid oscillations across the channel for an electron with the same initial conditions. As $\gamma \omega_p$ approaches $\omega$ with the increase of $a_0$, the time to traverse the channel becomes comparable to the period of the modulations of the restoring force. This leads to a continuous increase of the amplitude of the transverse oscillations shown in lower panel of Fig.~\ref{fig:traj}. The underlying cause becomes clear once we compare the electron trajectory with the modulation pattern. Each modulation effectively provides a kick towards the axis of the channel. Once the electron is kicked, it starts moving towards the axis, which means that the kick has occurred near the turning point. During the electron motion towards the axis, the restoring force drops because the time to traverse the channel is comparable to the period of the force modulations. The decrease of the force in combination with the extra momentum gained at the previous turning point allows the electron to travel further from the axis before experiencing the next kick. The situation repeats itself and each time the turning point moves further from the axis, indicating that the amplitude of the oscillations grows. 

This discussion emphasizes that the parametric amplification of betatron oscillations can be caused by non-inertial (accelerated/decelerated) relativistic axial motion even in the absence of a transverse oscillating field across the channel. The interplay between the laser field and the electrostatic field of the channel occurs by means of the the relativistic $\gamma$-factor that couples the transverse electron motion to the axial motion induced by the wave.

%


\begin{figure}[tb]
  \centering
  \subfigure{\includegraphics[scale=0.55]{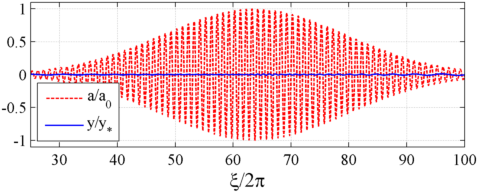} } \\
  \subfigure{\includegraphics[scale=0.55]{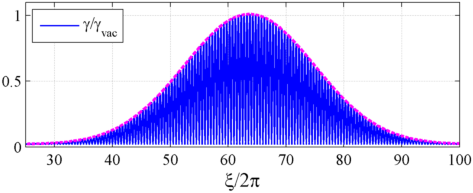}} 
\caption{Stable electron motion in the channel and the corresponding $\gamma$-factor at $\theta = 0$ and $\omega_p/\omega = 0.16$. The dashed (red) curve in the upper panel is the amplitude of the laser pulse sampled by the electron. The dashed (magenta) curve in the lower panel is the maximum $\gamma$-factor for the same pulse in the vacuum regime.}
\label{fig:S_stable}
\end{figure}

\section{Enhancement of laser-driven acceleration} \label{Sec_5}

In this section, we present examples of electron dynamics for different channel densities $n_0$ and polarizations angles $\theta$ that illustrate the key points of the qualitative discussion of Sec.~\ref{Sec_2}. We find the electron dynamics by solving Eqs.~(\ref{Eq1})~-~(\ref{Eq4}) numerically for a  pulse $a = a_*(\xi) \sin(\xi)$ with a Gaussian envelope
\begin{equation} \label{pulse_a}
a_* (\xi) = a_0 \exp\left[ - \frac{(\xi - \xi_0)^2}{2 \sigma^2} \right],
\end{equation}
where $a_0 = 10$, $\xi_0 = 400$, and $\sigma = 100$. The parameters $\xi_0$ and $\sigma$ determine the initial position of the laser beam with respect to the electron and the beam duration. In all the examples, the initial conditions for the electron at $\tau = 0$, which corresponds to $t = 0$, are the same: $p_y=0$, $p_z = p_0 = 0$, $y = y_0 = 0.05 c/\omega$, and $\xi = 0$, which corresponds to $z = 0$. The only two parameters that we vary from example to example are $n_0$ and $\theta$.


\begin{figure}[tb]
  \centering
  \subfigure{\includegraphics[scale=0.55]{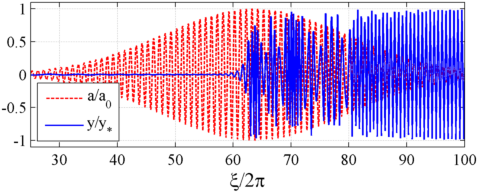} } \\
  \subfigure{\includegraphics[scale=0.55]{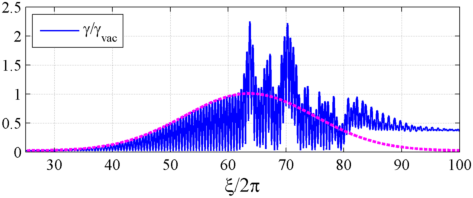}} 
\caption{Unstable electron oscillations across the channel and the corresponding $\gamma$-factor at $\theta = 0$ and $\omega_p/\omega = 0.175$. The dashed (red) curve in the upper panel is the amplitude of the laser pulse sampled by the electron. The dashed (magenta) curve in the lower panel is the maximum $\gamma$-factor for the same pulse in the vacuum regime.}
\label{fig:S_unstable}
\end{figure}

The upper panel of Fig.~\ref{fig:S_stable} shows the laser pulse and the amplitude of the transverse oscillations for $\theta = 0$. In this case, the laser electric field does not directly drive the transverse oscillations across the channel (along the $y$-axis), because it is directed along the $x$-axis. The low amplitude oscillations shown in Fig.~\ref{fig:S_stable} are caused by the initial displacement $y_0$. The ion density is set at $n_0 = 0.0256 n_{cr}$ ($\omega_p/\omega = 0.16$), so that the maximum amplitude of the transverse oscillations is $y_* \approx 8.1 c/\omega$ and $y_0/y_* \approx 5.7 \times 10^{-3}$. 

The laser pulse pushes the electron forward, causing the maximum $\gamma$-factor to increase with the laser envelope $a_*$ (see lower panel of Fig.~\ref{fig:S_stable}). The natural frequency of the transverse oscillations is proportional to $\sqrt{\gamma}$ [see Eq.~(\ref{main1})] and, therefore, it also increases with $a_*$. There is no exponential amplification of the transverse oscillations in Fig.~\ref{fig:S_stable}, which indicates that the natural frequency remains below the critical value needed for the parametric resonance during the entire pulse. According to Eq.~(\ref{thr2}), the ion density must exceed $0.026 n_{cr}$ for $a_0 = 10$, which is the maximum pulse amplitude, in order for the oscillations to become unstable. The ion density in our case is less than that. There is no visible enhancement of $\gamma$ compared to the vacuum regime. The dashed (magenta) curve in lower panel of Fig.~\ref{fig:S_stable} is the maximum $\gamma$-factor for the same pulse in the vacuum regime. It is calculated using Eq.~(\ref{main3_vac}) with $a=a_*$, where $a_*$ is defined by Eq.~(\ref{pulse_a}). The $\gamma$-factor for $n_0 = 0.0256 n_{cr}$ oscillates remaining below this curve.

\begin{figure}[tb]
  \centering
  \subfigure{\includegraphics[scale=0.55]{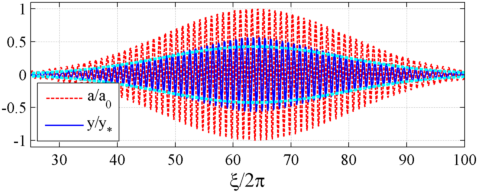} } \\
  \subfigure{\includegraphics[scale=0.55]{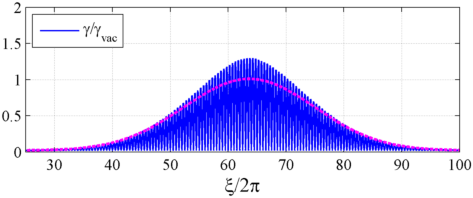}} 
\caption{Stable driven electron motion across the channel and the corresponding $\gamma$-factor at $\theta = \pi/4$ and $\omega_p/\omega = 0.07$. The dashed (red) curve in the upper panel is the amplitude of the laser pulse sampled by the electron. The dashed (magenta) curve in the lower panel is the maximum $\gamma$-factor for the same pulse in the vacuum regime.}
\label{fig:P_stable}
\end{figure}

Figure~\ref{fig:S_unstable} shows the electron dynamics at slightly higher density, $n_0 = 0.0306 n_{cr}$  $(\omega_p/\omega = 0.175)$, for the same laser polarization $(\theta = 0)$. At this density, the natural frequency of the transverse oscillations exceeds the critical value for $a_* > 9.3$ [see Eq.~(\ref{thr2})]. We therefore expect that the transverse electron oscillations should become unstable near the center of the pulse, since the maximum value of $a$ in the considered pulse is $a_0 = 10$. The transverse electron oscillations in Fig.~\ref{fig:S_unstable} indeed experience a rapid growth, which is in good agreement with the parametric mechanism described in Sec.~\ref{Sec_2_amp}. The oscillations are quickly amplified to the level comparable to $y_*$ and, at that point, the dephasing $d\xi/d \tau$ becomes significantly reduced. This allows for the electron to stay longer in phase with the wave, leading to a considerable enhancement of the laser-driven acceleration and a resulting enhancement of $\gamma$, as shown in the lower panel of Fig.~\ref{fig:S_unstable}. It is important to point out that the electron retains a significant portion of $\gamma_{vac}$ after being overtaken by the laser pulse. In contrast with that, the electron retains virtually no energy in the regime where the oscillations across the channel are stable (see Fig.~\ref{fig:S_stable}).

Next, we consider two regimes where the laser electric field across the channel is significant ($\theta = \pi/4$) and, therefore, it can directly drive betatron oscillations. Figure~\ref{fig:P_stable} shows the electron dynamics for $\theta = \pi/4$ and $n_0 \approx 0.007 n_{cr}$  $(\omega_p/\omega = 0.085)$. At this density, the maximum transverse displacement is $y_* \approx 16.6 c/\omega$, so that the normalized initial displacement is relatively small, $y_0/y_* \approx 3 \times 10^{-3}$. The amplitude of the transverse oscillations increases with the wave envelope $a_*$, because the oscillations are driven by the laser. Their amplitude becomes comparable to $y_*$ towards the middle of the pulse, with $y \approx y_*$ at $a_* = 10$ (see upper panel of Fig.~\ref{fig:P_stable}).

\begin{figure}[tb]
  \centering
  \subfigure{\includegraphics[scale=0.55]{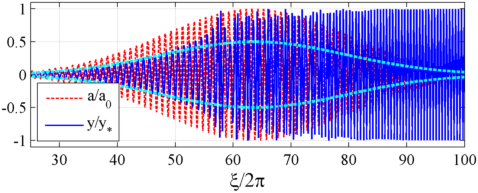} } \\
  \subfigure{\includegraphics[scale=0.55]{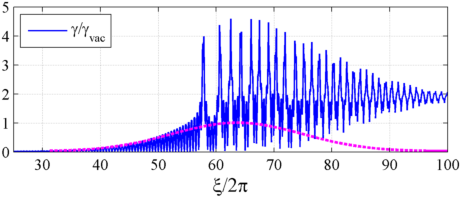}} 
\caption{Unstable electron oscillations across the channel and the corresponding $\gamma$-factor at $\theta = \pi/4$ and $\omega_p/\omega = 0.075$. The dashed (red) curve in the upper panel is the amplitude of the laser pulse sampled by the electron. The dashed (magenta) curve in the lower panel is the maximum $\gamma$-factor for the same pulse in the vacuum regime.}
\label{fig:P_unstable}
\end{figure}

In order to determine the effect of the channel, we have calculated $y$ and $\gamma$ for the same pulse assuming that the electron is moving in vacuum. This was done using the results of Sec.~\ref{Sec_vac}. The dashed (cyan) line in the upper panel of Fig.~\ref{fig:P_stable} is the extend of the vacuum oscillations determined from Eq.~(\ref{sol-1_0}). The dashed (magenta) curve in the lower panel of Fig.~\ref{fig:P_stable} is the maximum $\gamma$-factor in the vacuum regime determined from Eq.~(\ref{main3_vac}) with $a=a_*$, where $a_*$ is defined by Eq.~(\ref{pulse_a}). Figure~\ref{fig:P_stable} indicates that $y$ and $\gamma$ in the presence of ions start to deviate from the vacuum solution only when the amplitude of the driven electron oscillations becomes comparable to $y_*$, which agrees well with the estimates of Sec.~\ref{Sec_2}. 

According to Sec.~\ref{Sec_2}, we expect for the natural frequency of the transverse oscillations to become comparable to the frequency of its own modulations and to the frequency of the driving force at higher ion densities. Figure~\ref{fig:P_unstable} shows the electron dynamics for $\theta = \pi/4$ and $n_0 \approx 0.01 n_{cr}$  $(\omega_p/\omega = 0.1)$. The density increase from $n_0 \approx 0.085 n_{cr}$ to $n_0 \approx 0.01 n_{cr}$ leads to a significant change in electron dynamics due to the mentioned frequency match. The transverse oscillations in Fig.~\ref{fig:P_unstable} undergo amplification as the laser field approaches its peak amplitude. Afterwards, the amplitude $|y|$ no longer follows the behavior of the envelope $a_*$. It remains significant, $|y| \gg y_0$, even after the laser pulse overtakes the electron. The amplification leads to a sharp drop in the dephasing and, consequently, to a significant amplification of the axial acceleration. The resulting $\gamma$-factor spikes following the amplification, reaching $4 \gamma_{vac}$. Similarly to the regime shown in Fig.~\ref{fig:S_unstable}, the electron retains a significant portion of its maximum energy.

The pronounced threshold of the $\gamma$-factor enhancement for $\theta = 0$ and $\theta = \pi/4$ suggests that this is a generic feature independent of the polarization angle. In both cases, the enhancement is accompanied by considerable energy retention following the interaction with the beam.


\begin{figure}[tb]
  \centering
  \subfigure{\includegraphics[scale=0.55]{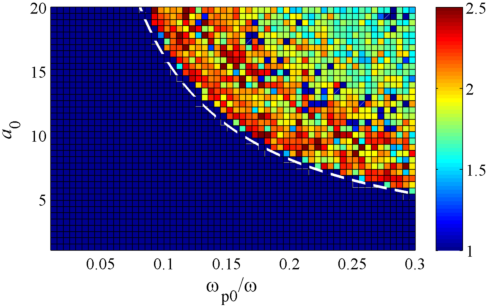} } 
  \subfigure{\includegraphics[scale=0.55]{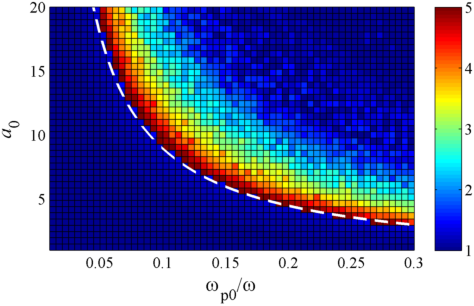}} 
\caption{Maximum $\gamma$-factor achieved by an electron irradiated by a pulse with maximum amplitude $a_0$ in a channel with plasma density $\omega_p$. The $\gamma$ is normalized to $\gamma_{vac} = 1 + a_0^2/2$ $(I \approx 1)$ defined by Eq.~(\ref{gamma_vac}). The upper and lower panels correspond to two different polarizations, $\theta=0$ and $\theta = \pi/4$.}
\label{fig:threshold_ex}
\end{figure}

\section{Threshold for energy enhancement} \label{Sec_6}

In Sec.~\ref{Sec_5}, we presented two examples of the enhancement of the axial acceleration that occurs with the increase of the maximum laser amplitude $a_0$ at a given ion density (or plasma frequency $\omega_p$) in the channel. In order to determine the extend of this effect, we have numerically solved Eqs.~(\ref{Eq1})~-~(\ref{Eq4}) for $1 \leq a_0 \leq 20$ and $0.01 \leq \omega_p \leq 0.3$. The laser pulse and  electron initial conditions are the same as in Sec.~\ref{Sec_5}. The only laser pulse parameter that we vary is the maximum amplitude $a_0$ of the envelope defined by Eq.~(\ref{pulse_a}). We determine the maximum $\gamma$-factor ($\max \gamma$) for each set of parameters from curves $\gamma(\xi)$ similar to those shown in Figs.~\ref{fig:S_stable} - \ref{fig:P_unstable}. 

The ratio $\max \gamma/\gamma_{vac}$ is shown for $\theta = 0$ and $\theta = \pi/4$ in Fig.~\ref{fig:threshold_ex}, where $\gamma_{vac} = 1 + a_0^2/2$ is the maximum $\gamma$-factor in the absence of ions. In both cases, there is a clear wave amplitude threshold for a given plasma frequency where $\max \gamma$ has a sharp jump. At $a_0 \geq 5$, the threshold for $\theta = 0$ (upper panel) agrees well with the parametric instability threshold $a_0 \approx 1.62 \omega/\omega_p$ that was derived in Sec.~\ref{Sec_2_amp} and that is shown with a dashed curve. Above the threshold, the enhancement factor $\max \gamma/\gamma_{vac}$ fluctuates between 1.5 and 3 for a wide range of parameters. The threshold for $\theta = \pi/4$ at $a_0 > 5$ is well matched by $a_0 \approx 0.89 \omega/\omega_p$ (dashed curve in the lower panel of Fig.~\ref{fig:threshold_ex}), which suggests that the threshold in this case is also determined by the combination $a_0 \omega_p/\omega$. The enhancement factor in this case also appears to be only a function of $a_0 \omega_p/\omega$.

\begin{figure}[tb]
  \centering
  \subfigure{\includegraphics[scale=0.45]{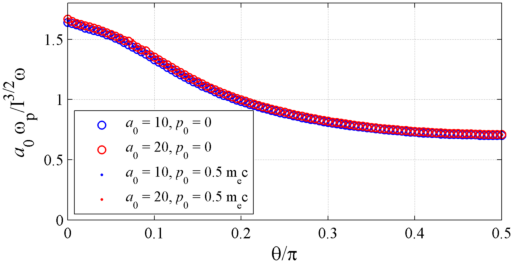} } 
  \subfigure{\includegraphics[scale=0.45]{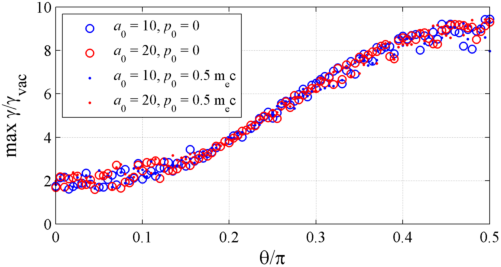}} 
\caption{Threshold for the enhancement of $\gamma$ as a function of the polarization angle. The  
upper and lower panels show $a_0 \omega_p/I^{3/2} \omega$ at the threshold and the enhancement factor $\gamma/\gamma_{vac}$ right above the threshold.}
\label{fig:threshold}
\end{figure}

After repeating the same parameter scan for polarization angles $0 \leq \theta \leq \pi/2$, we found that the sharp jump in $\max \gamma$ is present for all polarization angles. Circles in the upper panel of Fig.~\ref{fig:threshold} show $a_0 \omega_p /I^{3/2} \omega$ at the threshold as a function of $\theta$ for $a_0 = 10$ and $a_0=20$ for an electron without initial axial momentum ($p_0 = 0$, $I \approx 1$). In order to determine the role played by pre-acceleration, we have repeated the same procedure for an electron with initial axial momentum $p_0 = 0.5 m_e c$. The results are shown in Fig.~\ref{fig:threshold} with dots, which are not easily visible because they overlap with the circles. These plots clearly indicate that the onset of the electron energy enhancement is determined exclusively by the combination $a_0 \omega_p /I^{3/2} \omega$. This result is consistent with the estimate given by Eq.~(\ref{crit_amplitude}) in Sec.~\ref{Sec_2}. The dependence on $\theta$ is more subtle and because of that it is not captured by Eq.~(\ref{crit_amplitude}).

Our numerical results show that the value of $a_0 \omega_p /I^{3/2} \omega$ at the threshold is lowered by a factor of two with the increase of $\theta$ from 0 to $\pi/2$. This change in $\theta$ corresponds to a significant change in the amplitude of the driving force, because the component of the laser electric field across the channel increases from 0 to its maximum value. Therefore, we can conclude that the role of the driving electric field is secondary in determining the onset of the energy enhancement. 

The lower panel in Fig.~\ref{fig:threshold} shows by how much the $\gamma$ is enhanced right above the threshold compared to the $\gamma$ in the vacuum regime for the same wave amplitude ($\gamma_{vac}$). We find that the ratio $\max \gamma / \gamma_{vac}$ is not sensitive to the increase of $a_0$ or to the pre-acceleration. However, the enhancement factor $\max \gamma / \gamma_{vac}$ increases with the increase of $\theta$, which is correlated with the increase of the component of the laser electric field across the channel.
It must be pointed out that $\max \gamma$ does increase with the increase of $p_0$, but 
$\max \gamma / \gamma_{vac}$ remains unchanged because the maximum $\gamma$-factor that can be achieved in the vacuum regime ($\gamma_{vac}$) also increases with $p_0$ [see Eq.~(\ref{gamma_vac})].


\section{Electron Spectrum} \label{Sec_7}

We have so far discussed the energy enhancement and the corresponding threshold by considering a single electron. In this section, we consider an ensemble of test electrons with a spread in the initial transverse momentum and calculate the resulting time integrated spectrum for different channel densities and polarization angles.

\begin{figure}[tb]
  \centering
  \subfigure{\includegraphics[scale=0.5]{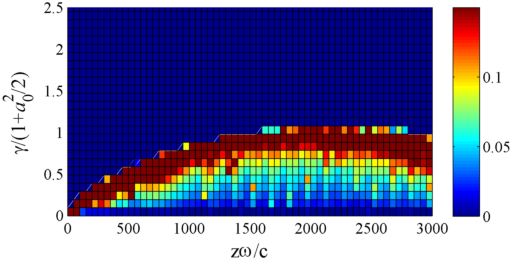} } \\
  \subfigure{\includegraphics[scale=0.5]{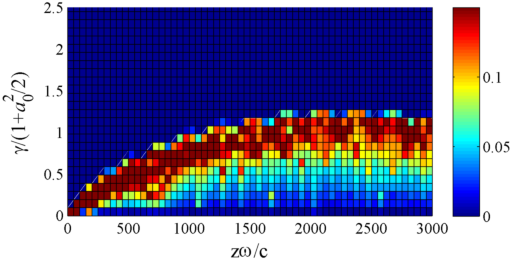}}  \\
  \subfigure{\includegraphics[scale=0.5]{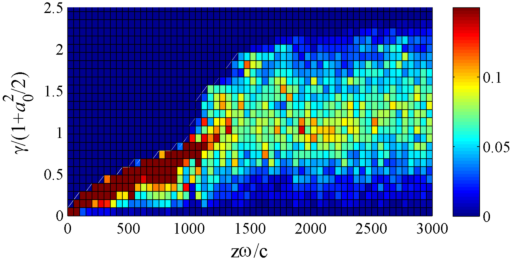} } \\
  \subfigure{\includegraphics[scale=0.5]{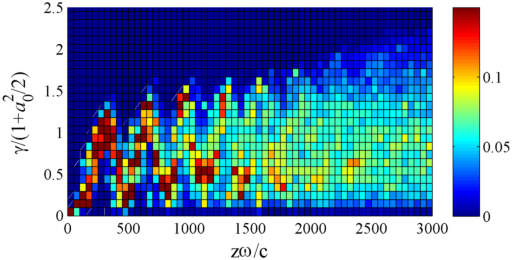}} 
\caption{Time-integrated electron spectrum as a function of distance traveled by an ensemble of electrons with a transverse momentum spread. The spectra from top to bottom are for: $\theta = 0$,  $\omega_p/\omega = 0.05$; $\theta = \pi/2$,  $\omega_p/\omega = 0.05$; $\theta = 0$,  $\omega_p/\omega = 0.2$; $\theta = \pi/2$,  $\omega_p/\omega = 0.2$. The color shows the relative fraction of electrons at a given axial location in a given energy bin.}
\label{fig:spectrum_comp}
\end{figure}

We consider electrons located initially at $z=y=0$ with an initial transverse momentum distribution $p_y = -p_y^{max},$ $-p_y^{max} + \Delta p_y$, ... $p_y^{max}$, where $p_y^{max} = 0.6 m_e c$, $\Delta p_y = 2p_y^{max}/(N_a + 1) = 1.2 \times 10^{-3} m_e c$, and $N_a = 1000$. The momentum spread is chosen such that the characteristic transverse momentum is relativistic, yet small compared to the maximum axial momentum in the vacuum regime. The integral of motion $I$ for this ensemble is in the range $1 < I < 1.17$. We use the same laser pulse as in Sec.~\ref{Sec_5}. For each electron, we find $\gamma(z)$ by first numerically solving Eqs.~(\ref{Eq1})~-~(\ref{Eq5}) to find $\gamma(\xi)$ and $t(\xi)$ and then by implicitly expressing $\xi$ in terms of $z$ using Eq.~(\ref{Eq_xi}). A histogram of electron energies for the entire ensemble at a given location $z$ gives a time-integrated electron spectrum that would be measured at this location. We split the $\gamma$-domain into bins with width $\Delta \gamma = 5$. 

Figure~\ref{fig:spectrum_comp} shows the electron spectra for four different sets of parameters, where the color indicates the relative fraction of electrons in a given energy bin at a given axial location. The top two panels correspond to $\omega_p/\omega = 0.05$ for $\theta = 0$ and $\theta = \pi/2$ and the bottom two panels correspond to $\omega_p/\omega = 0.2$ for the same polarizations. At $\omega_p/\omega = 0.05$, 
$a_0 \omega_p /I^{3/2} \omega$ is below the threshold for all electrons in the ensemble for both polarizations. In contrast with that, it is above the threshold at $\omega_p/\omega = 0.2$ for all the electrons (see Fig.~\ref{fig:threshold} for $a_0=10$). As a result, the electron spectra in the bottom two panels undergo enhancement and considerable broadening. 

The axial evolution of the electron spectra is determined by the wave envelope $a_* (\xi)$. The phase of the field sampled by each electron $\xi$ increases due to the continuous dephasing as the electron moves forward. Therefore, an electron has to travel a considerable axial distance before the wave envelope reaches its maximum. 
Below the amplification threshold ($\omega_p/\omega = 0.05$), the $\gamma$-factor follows the behavior of the envelope (for example, see Figs.~\ref{fig:S_stable} and \ref{fig:P_stable}). We can estimate the distance required for the $\gamma$-factor to reach its maximum value by taking into account that the electron is moving axially with the speed close to the speed of light, $z \approx c t$, and that the dephasing rate in this regime is $d \xi / d \tau \approx I$. We use Eq.~(8) to find that $t \approx \gamma \xi / \omega I$. Therefore, the axial distance traveled by the electron scales with the phase $\xi$ as $z \omega / c \approx \gamma \xi / I$. In our case, the spread in $I$ is relatively small, so all of the electrons will reach the maximum $\gamma$-factor close to the same axial location. The envelope reaches its maximum at $\xi = \xi_0$. We then take $I \approx 1$ and $\gamma \approx 50$ to find that the corresponding axial distance is $z \omega / c \approx 2000$. The spectra in 
the upper two panels of Fig.~\ref{fig:spectrum_comp} are consistent with this estimate. 

The estimate for $z$ as a function of $\xi$ also explains why the spectrum enhancement occurs at different axial locations for $\theta = 0$ and $\theta = \pi/2$ (the lower two panels of Fig.~\ref{fig:spectrum_comp}). The threshold condition for $\theta = \pi/2$ is satisfied at lower $a_*$ and correspondingly smaller $\xi$ than for $\theta = \pi/2$. Therefore, the electrons need to travel a shorter distance at $\theta = \pi/2$ before sampling the critical field. Note that the electron energy enhancement is comparable in both cases. As shown in Fig.~\ref{fig:threshold_ex}, the driving field across the channel is beneficial only if the combination $a_0 \omega_p / I^{3/2} \omega$ is relatively close to the threshold, whereas $a_0 \omega_p / I^{3/2} \omega$ is well above the threshold value for $\theta = \pi/2$ in our case.


\section{Summary and discussion} \label{sum}

We have analyzed the dynamics of an electron irradiated by a linearly polarized electromagnetic wave in a two-dimensional ion channel. It is shown that a considerable enhancement of the axial momentum and the total electron energy can be achieved via amplification of betatron oscillations. The equation for these oscillations is similar to that of an oscillator with a modulated natural frequency driven by an external force. The origin of the modulations is the non-inertial (accelerated/decelerated) relativistic axial motion induced by the wave. The frequency of the betatron oscillations increases with the wave amplitude and it is shown that the oscillations become parametrically unstable when their frequency becomes comparable to the frequency of the modulations. 

We have performed a parameter scan for a wide range of wave amplitudes $a_0$ and ion densities $n_0$ (or, equivalently, plasma frequencies $\omega_p$). We found that, for a given $n_0$, there is a well pronounced 
wave amplitude threshold above which the maximum electron energy is considerably enhanced. The threshold is determined by the ratio $a_0 \omega_p/I^{3/2} \omega$ whose value is only a function of the polarization angle $\theta$ (the angle between the laser electric field and the field of the channel), where $I$ is the integral of motion that accounts for possible pre-acceleration. We find that the enhancement factor $\gamma/\gamma_{vac}$ near the threshold is also only a function of $\theta$. We have calculated a time-integrated electron spectrum at various axial locations produced by an ensemble of electrons with a spread in the initial transverse momentum. The numerical results show that the energy enhancement is accompanied in this case by spectrum broadening.

In our analysis, we have treated the laser pulse as a plane wave propagating in vacuum. A self-consistent analysis 
that describes the channel formation is required to account for the effects of the channel on the wave, i.e., the change of the group and phase velocities. Another topic that requires attention is the mechanism of electron injection into the channel~\cite{Robinson2013}. The injection effectively determines the initial conditions for electron propagation in the channel and, therefore, it affects the threshold and the energy enhancement factor. 

\begin{figure}[tb]
  \centering
  \subfigure{\includegraphics[scale=0.55]{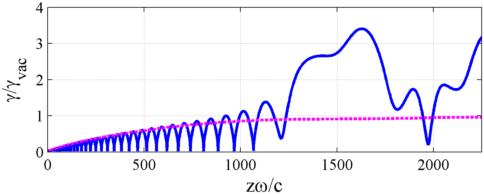} } \\
  \subfigure{\includegraphics[scale=0.55]{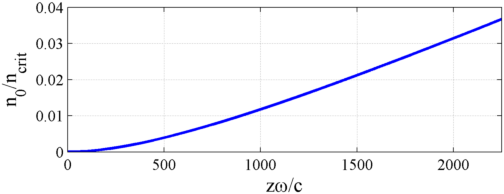}} 
\caption{The $\gamma$-factor of an electron irradiated by a laser pulse ($\theta = \pi/4$) in a channel with an axially increasing ion density. The pulse envelope is given by Eq.~(\ref{pulse_a}), where $a_0 = 10$, $\xi_0 = 400$, and $\sigma = 75$. The dashed (magenta) curve is the maximum $\gamma$-factor for the same pulse in the vacuum regime.}
\label{fig:den_example}
\end{figure}

In the case of laser-irradiated solid-density targets, the channel is formed in a preplasma that results from target expansion. Therefore, the ion density in the channel might be axially increasing. We however expect for the mechanism of energy enhancement to be robust with respect to the axial variations, because essentially it is a threshold phenomenon rather than a phenomenon that relies on a narrow linear resonance. To demonstrate that, we numerically solved Eqs.~(\ref{Eq1})~-~(\ref{Eq4}) for an ion density profile shown in Fig.~\ref{fig:den_example}. The resulting $\gamma$-factor exhibits a significant enhancement compared to the vacuum case despite a considerable density increase. The peaks of enhanced $\gamma$ are relatively wide, which can make it easier to generate hot electrons in a target using a preplasma. The electrons would be injected into the target for a wide range of preplasma lengths.

\section{Acknowledgments}

This work was supported by Sandia National Laboratory Contract No. PO 990947, National Nuclear Security Administration Contract No. DE-FC52-08NA28512, and U.S. Department of Energy Contract No. DE-FG02-04ER54742. Sandia National Laboratories is a multi-program laboratory managed and operated by Sandia Corporation, a wholly owned subsidiary of Lockheed Martin Corporation, for the U.S. Department of Energy’s National Nuclear Security Administration under contract DE-AC04-94AL85000.


\appendix
\section{Equations of motion for a test electron } \label{Ap-1}

The Hamiltonian for an electron irradiated by an electromagnetic wave with vector potential ${\bf{A}}$ in a steady-state ion channel with electrostatic potential $\varphi$  is  
\begin{equation}
	H = \sqrt{m_e^2 c^4 + c^2 \left( {\bf{P}} + |e| {\bf{A}} /c \right)^2} - |e| \varphi,
\end{equation}
where $e$ and $m_e$ are the electron charge and mass, $c$ is the speed of light, and ${\bf{P}}$ is the generalized momentum. The input functions ${\bf{A}}$ and $\varphi$ specify the polarization and shape of the laser beam and the geometry of the ion channel. We consider a straight channel, such that the corresponding electrostatic potential $\varphi$ depends only on transverse coordinates $x$ and $y$ and it is independent of the axial coordinate $z$. We assume that the incoming wave is a plane wave propagating along the axis of the channel, so that ${\bf{A}} = {\bf{A}} (z-ct)$.  

A point canonical transformation 
\begin{equation}
q = z - ct
\end{equation}
yields a new Hamiltionian $H' = H - c P_z$ that no longer explicitly depends on time, since $\varphi$ is time independent and ${\bf{A}}$ can be written only as a function of $q$. Therefore, $H' = H - c P_z$ is an integral of motion, i.e., $d H' / dt = 0$. The Hamilton's equations for $\varphi = \varphi({\bf{r}}_{\perp})$ and ${\bf{A}} = {\bf{A}} (q)$ have the form 
\begin{eqnarray}
&& \frac{d {\bf{P}}_{\perp}}{d t} = |e| \frac{\partial \varphi}{\partial {\bf{r}}_{\perp}}, \label{P_perp} \\
&& \frac{d P_z}{d t}  = -  \frac{c}{\gamma} \left( {\bf{P}}_{\perp} + m_e c {\bf{a}} \right) \frac{\partial {\bf{a}} }{\partial q}, \label{Ap_Pz}\\
&& \frac{d {\bf{r}}_{\perp}}{dt} = \frac{c}{\gamma} \left( {\bf{P}} / m_e c + {\bf{a}}  \right), \\
&& \frac{d q}{dt} = \frac{c}{\gamma} \left( P_z / m_e c - \gamma \right), \label{dq/dt}
\end{eqnarray}
where ${\bf{a}} \equiv |e| {\bf{A}} / m_e c^2$ is a dimensionless vector potential and
\begin{equation}
\gamma = \sqrt{1 + \left( {\bf{a}} + {\bf{P}}/m_e c\right)^2}
\end{equation}
is the relativistic factor. It might be convenient to replace Eq.~(\ref{Ap_Pz}) that describes the evolution of the axial momentum with the equation $d H' / dt = 0$ that can be written as
\begin{eqnarray}
&& \frac{d}{dt}  \left( \gamma m_e c^2 - |e| \varphi - c P_z \right)  =0. \label{int_motion_Ap}
\end{eqnarray}


In the case of a slab-like ion channel shown in Fig.~\ref{Fig3}, the electrostatic potential  is 
\begin{equation} \label{slab_Ey}
	\varphi = -2 \pi n_0 |e| y^2, 
\end{equation}
where $n_0$ is the density of the singly charged ions in the channel. The $x$-component of Eq.~(\ref{P_perp})
can be readily integrated to find that
\begin{equation}
	P_x = 0
\end{equation}
for an electron that is initially not moving along the $x$ axis $(P_x^{t = 0} = 0)$. The equations that describe the electron motion in the $(y,z)$ plane then take the form:
\begin{eqnarray}
&& \frac{d P_y}{d t} = - m_e \omega_p^2 y , \label{P_perp2} \\
&& \frac{d P_z}{d t}  = -  \frac{c}{\gamma} \left[ P_y \frac{\partial a_y}{\partial q} + m_e c \frac{\partial}{\partial q} \left( \frac{|{\bf{a}}|^2}{2} \right) \right] , \label{Ap_Pz2}\\
&& \frac{d y}{dt} = \frac{c}{\gamma} \left( P_y / m_e c + a_y  \right), \\
&& \frac{d q}{dt} = \frac{c}{\gamma} \left( P_z / m_e c - \gamma \right), \label{Ap_q}
\end{eqnarray}
where $\omega_p \equiv \sqrt{4 \pi n_0 e^2 / m_e}$. It follows from Eq.~(\ref{int_motion_Ap}) that the integral of motion for an electron moving in this channel is
\begin{eqnarray}
&& I \equiv \gamma + \frac{\omega_p^2}{c^2} \frac{y^2}{2} - \frac{P_z}{m_e c} . \label{R}
\end{eqnarray}

It is convenient to introduce a dimensionless proper time $\tau$ defined by the relation 
\begin{equation} \label{t-tau}
	d \tau /d t = \omega / \gamma,
\end{equation}
and a dimensionless phase variable 
\begin{equation} \label{s_Ap}
	\xi \equiv \omega (t - z / c) = - \omega q / c,
\end{equation}
where $\omega$ is the frequency of the electromagnetic wave. Equations~(\ref{P_perp2})~-~(\ref{Ap_q}) now take the form
\begin{eqnarray}
&& \frac{d }{d \tau} \left( \frac{P_y}{m_e c} \right) = - \gamma \frac{\omega_p^2}{\omega^2}  \frac{\omega}{c} y, \label{P_perp3} \\
&& \frac{d}{d \tau} \left( \frac{P_z}{m_e c} \right)  = \frac{P_y}{m_e c} \frac{\partial a_y}{\partial \xi} + \frac{\partial}{\partial \xi} \left( \frac{|{\bf{a}}|^2}{2} \right) , \label{Ap_Pz2}\\
&& \frac{d}{d \tau} \left( \frac{\omega}{c} y \right) =  \frac{P_y}{m_e c} + a_y , \label{Ap_Py2}\\
&& \frac{d \xi}{d \tau} =  \gamma - \frac{P_z}{m_e c}, \label{xi_Ap2}
\end{eqnarray}
where
\begin{equation}
\gamma = \sqrt{1 + a_x^2 + \left(a_y + P_y/m_e c\right)^2 + \left(P_z / m_e c \right)^2}. \label{gamma_2}
\end{equation}
Equations~(\ref{P_perp3}) - (\ref{xi_Ap2}) form a closed set of equations for the electron motion in the $(y,z)$-plane. One of the equations can be replaced with Eq.~(\ref{R}), which would give an equivalent closed set of equations.

\end{document}